\documentclass{iaus}
\usepackage{graphicx}

\title[The 6dF Galaxy Survey]
{The 6dF Galaxy Survey: a low-redshift benchmark for bulge-dominated galaxies}

\author[Colless et al.]
{Matthew Colless$^1$, Heath Jones$^1$, Rob Proctor$^2$, Craig
  Harrison$^{1,3}$, Lachlan Campbell$^{1,3}$ \and Philip Lah$^3$}

\affiliation{
$^1$Anglo-Australian Observatory, Epping, NSW, Australia\break 
    email: colless@aao.gov.au\\[\affilskip]
$^2$Centre for Astrophysics and Supercomputing, Swinburne University of
    Technology,\break Hawthorn, VIC, Australia\\[\affilskip]
$^3$Research School of Astronomy and Astrophysics, Australian National
    University,\break Weston Creek, ACT, Australia}

\pubyear{2007}
\volume{245}
\pagerange{1--4}
\date{?? and in revised form ??}
\jname{Formation and Evolution of Galaxy Bulges}
\editors{Martin Bureau, Lia Athanassoula, Beatriz Barbuy eds}
\begin{document}

\maketitle

\begin{abstract}

  The 6dF Galaxy Survey provides a very large sample of galaxies with
  reliable measurements of Lick line indices and velocity dispersions.
  This sample can be used to explore the correlations between mass and
  stellar population parameters such as age, metallicity and
  [$\alpha$/Fe]. Preliminary results from such an analysis are presented
  here, and show that age and metallicity are significantly
  anti-correlated for both passive and star-forming galaxies. Passive
  galaxies have strong correlations between mass and metallicity and
  between age and $\alpha$-element over-abundance, which combine to
  produce a downsizing relation between age and mass. For old passive
  galaxies, the different trends of $M/L$ with mass and luminosity in
  different passbands result from the differential effect of the
  mass--metallicity relation on the luminosities in each passband.
  Future work with this sample will examine the Fundamental Plane of
  bulge-dominated galaxies and the influence of environment on relations
  between stellar population parameters and mass.


\end{abstract}

\section{The 6dF Galaxy Survey}

The 6dF Galaxy Survey (6dFGS; www.aao.gov.au/local/www/6df; Jones
et~al.\ 2004, 2005) is a redshift and peculiar velocity survey of
galaxies in the local universe. The observations for the survey were
obtained during 2001--2006 using the UK Schmidt Telescope and the 6dF
spectrograph (Watson et~al.\ 1998). The 6dFGS covers 92\% of the
southern sky with $|b|>10^\circ$. Its primary sample is drawn from the
2MASS Extended Source Catalog (XSC; Jarrett et~al.\ 2000) and consists
of galaxies with $K_{\rm tot}<12.65$; for this sample the redshift
completeness is 88\%. It also includes secondary samples complete down
to $H<12.95$, $J<13.75$ (from the 2MASS XSC) and $r_F<15.6$, $b_J<16.75$
(from the SuperCosmos Sky Survey; Hambly et~al.\ 2001). The 6dFGS
peculiar velocity survey uses the Fundamental Plane relation to derive
distances and velocities for about 15,000 bright early-type galaxies.
The 6dFGS database comprises 137k spectra and 124k galaxy redshifts,
plus photometry and images. The final data release will be made public
in August 2007 (Jones et~al.\ in prep.; see www-wfau.roe.ac.uk/6dfgs).

As well as its intended purpose as a survey of the structure and motions
in the local universe, the 6dFGS also provides a benchmark sample for
studying the properties of the low-redshift galaxy population. Here we
present some preliminary results from an analysis of the stellar
populations in a sample of about 6000 galaxies with high-quality spectra
from the 6dFGS Second Data Release (DR2; Jones et~al.\ 2005). Velocity
dispersions ($\sigma$) have been measured for these galaxies using the
cross-correlation technique of Tonry \& Davis (1979). Comparisons with
other high-quality samples show good agreement and imply the 6dFGS
dispersions have a median error of 10.9\% (Campbell et~al., in prep.).

\section{Stellar population parameters and relations}

Stellar population parameters for these galaxies have been derived from
Lick index measurements for approximately 15 indices covering the
spectral range 4000--5500\AA\ (Proctor et~al., in prep.). Ages,
metallicities and $\alpha$-element over-abundances were determined by
comparing these index measurements to the simple stellar population
models of Korn, Maraston \& Thomas (2005) using the $\chi^2$-fitting
method introduced by Proctor \& Sansom (2002). The method rejects those
indices that give poor fits to the models, and this excludes H$\beta$
for galaxies both with and without H$\alpha$ emission. For passive
galaxies without H$\alpha$ emission, the uncertainties in the derived
ages are typically about 0.17~dex in age, 0.15~dex in [Z/H] and 0.10~dex
in [$\alpha$/Fe]. For galaxies that do show significant H$\alpha$
emission, the fits also reject H$\gamma$ and H$\delta$, which results in
somewhat larger uncertainties.

\begin{figure}
 \centerline{
 \includegraphics[angle=270, width=\textwidth]{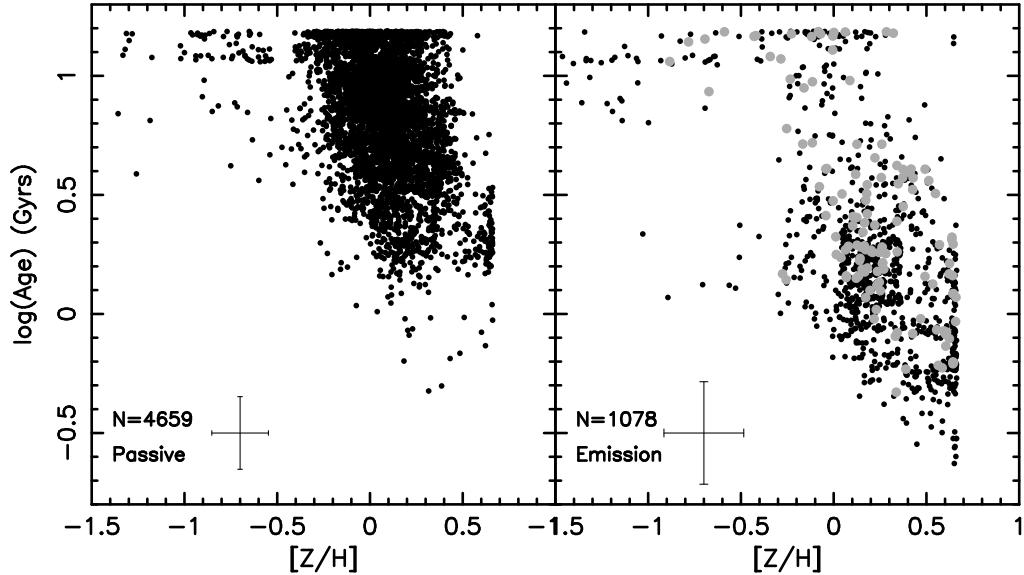}}
 \caption{The distribution of age and metallicity for the passive
   galaxies (left) and the emission-line galaxies (right). Galaxies
   classified as AGN on the basis of their emission line ratios are
   shown as grey dots in the right-hand panel. The mean uncertainties in
   the estimates of age and metallicity are shown by the error bars in
   the bottom left corner of each panel.}\label{fig:agez}
\end{figure}

The distribution of age and metallicity is shown in
Figure~\ref{fig:agez} both for passive galaxies (those with no detected
H$\alpha$ emission) and for star-forming galaxies (those in which
H$\alpha$ emission is detected). In general, the passive galaxies have
metallicities in the range [Z/H]=$-0.3$--$0.5$ and luminosity-weighted
ages of 3--14\,Gyr, while the star-forming galaxies have solar or higher
metallicities and ages less than 3\,Gyr. Both samples show a trend of
increasing metallicity for younger galaxies that is substantially
stronger than the residual correlation between the errors in age and
metallicity.

\begin{figure}
 \centerline{
 \includegraphics[angle=0, width=\textwidth]{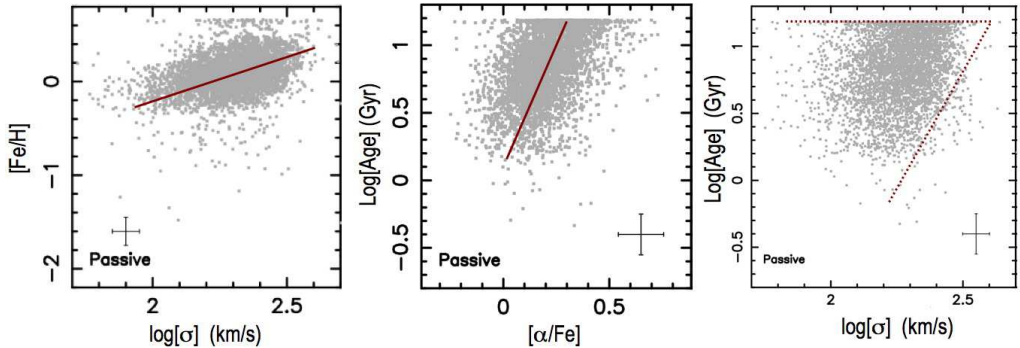}}
 \caption{For passive galaxies there are strong correlations between
   metallicity and velocity dispersion (left panel) and between age and
   [$alpha$/Fe] (middle panel); there is also a downsizing relation
   between age and velocity dispersion (right panel).}\label{fig:trends}
\end{figure}

If we just consider the passive galaxies, then as well as the
correlation between age and metallicity, there are also correlations
between metallicity and velocity dispersion and between age and
[$\alpha$/Fe], as shown in the left and middle panels of
Figure~\ref{fig:trends}: more massive galaxies tend to have higher
metallicities, while older galaxies tend to have larger $\alpha$-element
over-abundances. There is also a downsizing relation between age and
velocity dispersion, in which the age of the {\it youngest} galaxies
decreases with decreasing velocity dispersion, as indicated in the right
panel of Figure~\ref{fig:trends}. This latter relation would appear to
be the projection in the age--$\log\,\sigma$ plane of the
[Z/H]--$\log\,\sigma$ and age--[$\alpha$/Fe] relations.

\section{Age and metallicity effects on $M/L$}

\begin{figure}
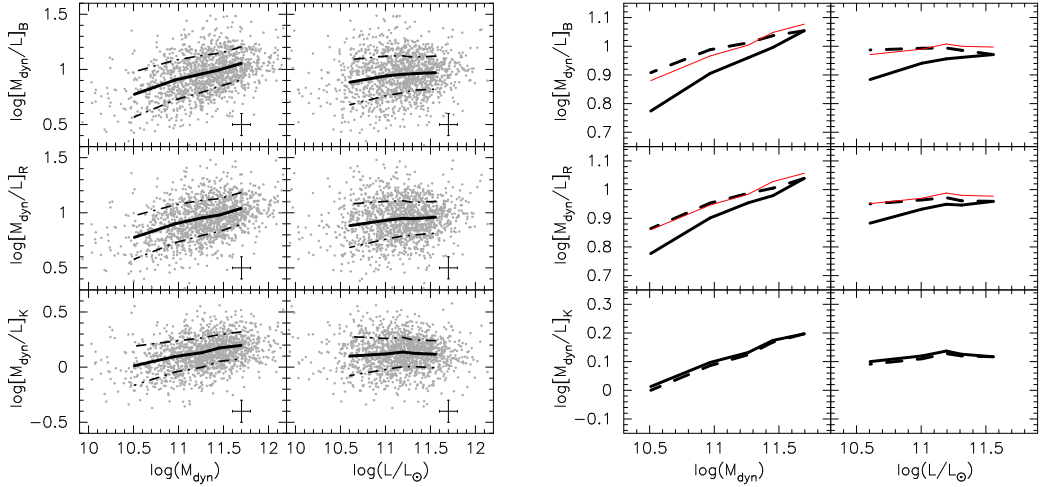

 \centerline{
 \includegraphics[angle=270, width=0.48\textwidth]{colless_iau245_fig3a.eps}
 \hspace{0.04\textwidth}
 \includegraphics[angle=270, width=0.48\textwidth]{colless_iau245_fig3b.eps}}
\caption{At left, the dynamical mass-to-light ($M/L$) ratios of old
  galaxies (age $>$10\,Gyr) in $B$, $R$ and $K$ bands is plotted against
  dynamical mass and $K$-band luminosity. Solid lines are means in five
  bins along the $x$-axis; the rms scatter is indicated by dashed lines.
  The average error on individual points is shown at the bottom right of
  each panel. At right, the mean $M/L$ ratios are shown before (solid
  line) and after (dashed line) correction for the effect of the
  mass--metallicity relation on the luminosity. The thin solid lines in
  each panel show the $K$-band relation for the purposes of
  comparison.}\label{fig:mtol}
\end{figure}

It is interesting to examine the effects of age and metallicity on the
mass-to-light ratios for galaxies in different passbands. If we consider
only old galaxies (greater than 10\,Gyr), then the left panel of
Figure~\ref{fig:mtol} shows the trends of $M/L$ in the $B$, $R$ and $K$
bands against both dynamical mass and $K$-band luminosity. Two points
are worth noting: (i)~the trends with dynamical mass are steeper than
those with luminosity in all passbands, and (ii)~the trends with both
mass and luminosity are steeper in bluer passbands.

As well as eliminating age effects by selecting only old galaxies, we can
also attempt to remove the dependence on metallicity in these relations.
The right panel of Figure~\ref{fig:mtol} shows as thick solid lines the
same binned relations as in the left panel; it also shows these
relations after correcting the mass-to-light ratio for differences in
metallicity (dashed lines). These corrections are determined from the
luminosities predicted by Bruzual \& Charlot (2003) models for stellar
populations having the metallicity (based on the mass--metallicity
relation) and age of each mass or luminosity bin, and converting to the
predicted luminosity for a stellar population with the same metallicity
as the highest mass or luminosity bin. The corrected mass-to-light
ratios resulting from this procedure have almost identical dependences
on mass or luminosity, regardless of passband, as can be seen by
comparing the dashed lines in the right panel of Figure~\ref{fig:mtol}
with the thin solid lines (which are just the corrected relations for
the $K$-band). Thus the variations of $M/L$ with passband for this
sample of old galaxies can be fully accounted for by the
mass--metallicity relation. It is worth noting that the correction has
almost no effect in the $K$-band, underlining the minimal impact of
metallicity on $K$-band luminosities and the usefulness of this passband
for studies of $M/L$ and the Fundamental Plane.

\section{Conclusions}

The 6dFGS provides a large sample of galaxies with spectra of
sufficiently high quality and spectral range to allow reliable
measurements of both line indices and velocity dispersions. This sample
can be used to explore the correlations between mass, age, metallicity
and [$\alpha$/Fe]. Some preliminary results from such an analysis are
presented here; a more detailed study will appear in Proctor et~al.\
(2007, in prep). 

We find that age and metallicity are significantly anti-correlated
(younger galaxies are more metal-rich) for both passive and star-forming
samples. Passive galaxies show the well-known strong correlation between
mass and metallicity (more massive galaxies are more metal-rich), but
also a strong correlation between age and $\alpha$-element
over-abundance (older galaxies have higher over-abundances). These two
effects combine to produce a downsizing relation between age and mass,
such that the age of the {\it youngest} galaxies decreases with
decreasing mass. For old ($>$10\,Gyr) passive galaxies, the different
trends of $M/L$ with mass and luminosity in different passbands result
from the differential effect of the mass--metallicity relation on the
luminosities in each passband.

Future work will examine the Fundamental Plane of bulge-dominated
galaxies and the influence of environment on relations between stellar
population parameters and mass.

\begin{acknowledgments}
  We acknowledge the contributions to this research program of the UKST
  observers and the entire 6dFGS team
  (www.aao.gov.au/local/www/6df/6dFGSteam.html).
\end{acknowledgments}


\begin{thebibliography}{}

\bibitem[Bruzual \& Charlot (2003)]{Bruzual03}
     {Bruzual G., Charlot S.}, 2003,
     \textit{MNRAS}, 344, 1000

\bibitem[Hambly et~al.\ (2001)]{Hambly01}
     {Hambly N.C., MacGillivray H.T., Read M.A., et~al.}, 2001,
     \textit{MNRAS}, 326, 1295

\bibitem[Jarrett et~al.\ (2000)]{Jarrett00}
     {Jarrett T.H., Chester T., Cutri R., et~al.}, 2000,
     \textit{AJ}, 119, 2498

\bibitem[Jones et~al.\ (2004)]{Jones04}
     {Jones D.H., Saunders W., Colless M., et~al.}, 2004,
     \textit{MNRAS}, 355, 747

\bibitem[Jones et~al.\ (2005)]{Jones05}
     {Jones D.H., Saunders W., Read M., Colless M.}, 2005,
     \textit{PASA}, 22, 277

\bibitem[Korn et~al.\ (2005)]{Korn05}
     {Korn A.J., Maraston C., Thomas D.}, 2005,
     \textit{A\&A},438, 685

\bibitem[Proctor \& Sansom (2002)]{Proctor02}
     {Proctor R.N., Sansom A.E.}, 1998,
     \textit{SPIE}, 3355, 834

\bibitem[Watson et~al.\ (1998)]{Watson98}
     {Watson F.G., Parker Q.A., Miziarski S.}, 1998,
     \textit{SPIE}, 3355, 834

\end{thebibliography}
\end{document}